\documentclass{pnastwo}
\usepackage[numbers,round]{natbib}
\usepackage{hyperref}
\hypersetup{
    pdfcreator={Hello},
    pdfproducer={World}
}

\begin{document}

\title{Jupiter's Decisive Role in the Inner Solar System's Early Evolution}

\author{Konstantin Batygin\affil{1}{Division of Geological and Planetary Sciences, California Institute of Technology, 1200 E. California Blvd., Pasadena, CA 91125, USA} \and Gregory Laughlin\affil{2}{Department of Astronomy \& Astrophysics, UCO/Lick Observatory, University of California, Santa Cruz, Santa Cruz, CA 95064, USA}}

\contributor{Submitted to Proceedings of the National Academy of Sciences of the United States of America}


\maketitle

\begin{article}
\begin{abstract}
{The statistics of extrasolar planetary systems indicate that the default mode of planet formation generates planets with orbital periods shorter than $100$ days, and masses substantially exceeding that of the Earth. When viewed in this context, the Solar System is unusual. Here, we present simulations which show that a popular formation scenario for Jupiter and Saturn, in which Jupiter migrates inward from $a>5\, \rm{AU}$ to $a\sim1.5\,\rm{AU}$ before reversing direction, can explain the low overall mass of the Solar System's terrestrial planets, as well as the absence of planets with $a<0.4\,\rm{AU}$. Jupiter's inward migration entrained $s\gtrsim 10-100\,\rm{km}$ planetesimals into low-order mean-motion resonances, shepherding and exciting their orbits. The resulting collisional cascade generated a planetesimal disk that, evolving under gas drag, would have driven any pre-existing short-period planets into the Sun. In this scenario, the Solar System's terrestrial planets formed from gas-starved mass-depleted debris that remained after the primary period of dynamical evolution.}
\end{abstract}



\section{Significance}
\dropcap{T}he Solar System is an unusual member of the galactic planetary census in that it lacks planets that reside in close proximity to the Sun. In this work, we propose that the primordial nebula-driven process responsible for retention of Jupiter and Saturn at large orbital radii and sculpting Mars' low mass is also responsible for clearing out the Solar System's innermost region. Cumulatively, our results place the Solar System and the mechanisms that shaped its unique orbital architecture into a broader, extrasolar context.

\section{Introduction}
A full understanding of the formation and the early evolution of the Solar System ranks among natural science's grand challenges, and at present, even the dominant processes responsible for generating the observed planetary architecture remain elusive. Nonetheless, the past three decades have generated remarkable progress \citep{Morby2012}, and critically, the discovery of thousands of extrasolar planets has placed the Earth and the Solar System into the broader context of the galactic planetary census.

Perhaps the most important exoplanet-related discovery  has been the realization that roughly half of the Sun-like stars in the solar neighborhood are accompanied by systems of one or more planets on low-eccentricity orbits with periods ranging from days to months, and masses falling in the $1\,$M$_{\oplus}<M_{\rm p}<50\,$M$_{\oplus}$ range \citep{Mayor2011,Batalha2013}. This dominant population of planets (which often presents tightly packed, nearly co-planar multiple systems) contrasts sharply with the Solar System, whose inner edge is marked by Mercury's 88-day (0.4 AU) orbit (see Figure 1). An iconic example from the new planetary catalog is the \textit{Kepler-}11 system, which encompasses at least six planets comprising more than $\sim40$ Earth masses \citep{Lissauer2011_1}. In short, the exoplanetary surveys have revealed a hitherto unrecognized oddity of the Solar System. Relative to other Sun-like, planet-bearing stars, our terrestrial region is severely depleted in mass.

A few related peculiarities are also evident within the inner Solar System. Specifically, cosmo-chemical evidence suggests that while the fundamental planetary building blocks (planetesimals) formed within $\sim1$ Myr of the Sun's birth \citep{Connelly2012}, the final assembly of the terrestrial planets occurred on a timescale of $100-200$ Myr, well after the dispersal of the nebular gas \citep{Chambers2011}. This is at odds with the inferred compositions of extrasolar Super-Earths, which are thought to have substantial gaseous atmospheres. Additionally, the exceptionally small masses of Mercury and Mars suggest\footnote{See refs \citep{FischerCiesla2014,Izidoro2014} for an alternative view.} that the terrestrial planets formed out of a narrow annulus of rocky debris, spanning $0.7-1$ AU (where $1$AU is the mean distance between the Earth and the Sun) \citep{Hansen2009}. Such a narrow annulus is at odds with so-called minimum mass Solar Nebula \citep{Hayashi1981,Desch2007}. 

Within the framework of a radially confined solid component of the inner Solar nebula, the inner edge of the annulus is entirely artificial. Indeed, at present there exists no compelling justification for its origin. A plausible explanation may stem from the dynamical evacuation of solid material by a population of primordial close-in planets \citep{Boley2014}. We shall investigate this possibility further in this study.

Unlike the inner edge of the annulus, a body of previous work has demonstrated that the outer edge can be naturally sculpted by inward-then-outward migration of Jupiter \citep{Walsh2011_1}. Within protoplanetary disks, long-range migration of giant planets results from tidal interactions with the nebula and viscous transport \citep{KleyNelson2012}. For single planets, orbital evolution is typically inwards. However, the process of resonant locking between two convergently migrating planets can lead to a reversal of the migration direction \citep{MassetSnellgrove2001}.

\begin{figure}[t]
\label{fig1}
\includegraphics[width=0.95\columnwidth]{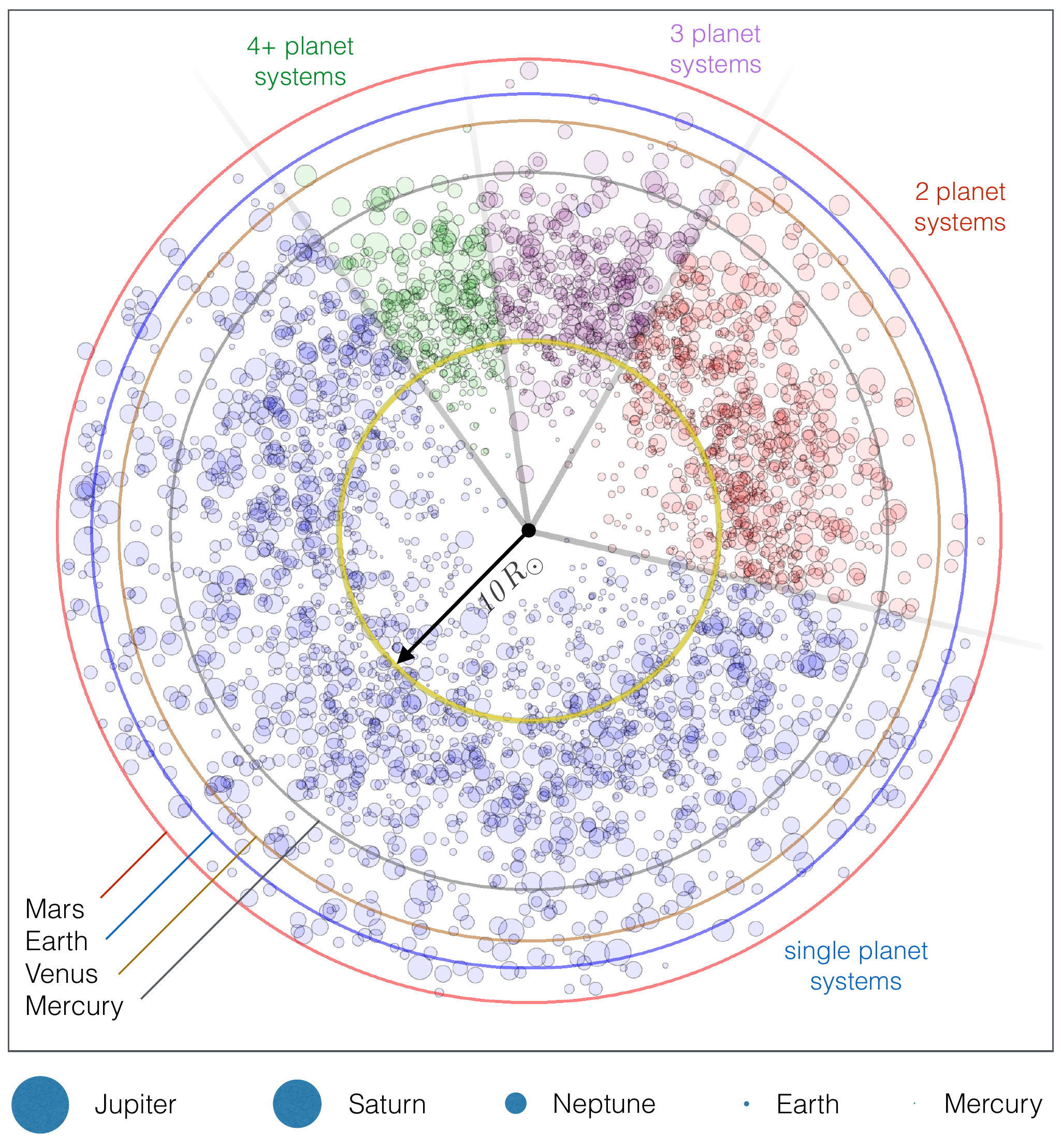}
\caption{\small{Orbital distribution of sub-Jovian extrasolar planets. A collection of transiting planet candidates with radii $R < 5 R_{\oplus}$, detected by the \textit{Kepler} mission is shown. The radial distance away from the center of the figure represents a logarithmic measure of the planetary semi-major axis, such that the origin corresponds to the Sun's surface. The sizes of the individual points represent the physical radii of the planets. Further, the points are color-coded in accordance with multiplicity. The orbits of the terrestrial planets are also shown. Despite observational biases inherent to the observed distribution (e.g. transit probability, detectability) that work against detection of planets at increasing orbital radii, the raw contrast to our own Solar System is striking.}} 
\end{figure}

The process of resonant migration reversal for gap-opening planets (i.e. objects with $M\gtrsim M_{\rm{Jup}}$) is a well-understood result of planet-disk interactions, and only requires the outer planet to be somewhat less massive than the inner. To this end, it is worth noting that all of the known mean-motion commensurate pairs of giant planets that reside beyond $a\gtrsim1\, \rm{AU}$ have the more massive object on the inside \citep{Wright2011,Morbidelli2013}, suggesting that the operation of this mechanism is widespread\footnote{A notable system within the resonant extrasolar population is GJ 876, where the inner planet is substantially less massive than the outer. In accordance with the picture of resonant transport delineated in ref. \citep{MorbidelliCrida2007}, this system likely failed to satisfy the conditions required for migration reversal and decayed to a compact orbital configuration \citep{LeePeale2002}.}.

Within the Solar System, it is inferred that Jupiter initially migrated inwards from its primordial formation site (presumably $3-10$ AU) to $\sim 1.5$ AU, and subsequently reversed its evolutionary track as a consequence of locking into a 3:2 mean-motion resonance with a newly-formed Saturn. This special case of the generic resonant migration reversal mechanism is informally referred to as the ``Grand Tack" scenario \citep{Walsh2011_1}. In addition to the aforementioned truncation of the inner solid nebula, this putative sequence of events is attractive in that it naturally explains how the Solar System's giant planets avoided spiraling into Sun \citep{MorbidelliCrida2007}, accounts for the origins of compositional differences within the Asteroid belt \citep{Walsh2011_1}, provides a mechanism for delivery of water into the terrestrial region \citep{OBrien2014}, and generates a compact orbital configuration needed for the subsequent instability-driven orbital evolution of the outer Solar System \citep{Morby2007,BatyginBrown2010}. 

\section{Resonant Transport and Collisional Evolution}
An early inward migration for Jupiter has a number of repercussions that come to light when one places the Solar System into the broader context provided by the observations of extrasolar planets. An inescapable consequence of Jupiter's trek is the resonant capture and the ensuing inward entrainment and transport of solid material \citep{Malhotra1993_1}. In particular, when a given planetesimal's orbital period becomes nearly rational multiple of the orbital period of Jupiter, gravitational perturbations become coherent and force the planetesimal to maintain the same period ratio, leading to a decrease in the planetesimal orbit's semi-major axis \citep{YuTrem2001_1}. The most common commensurability at which capture occurs is 2:1, although numerous other possibilities exist.

In order for resonant interactions to be effective, the planetesimal in question must not be hydrodynamically coupled to the nebular gas on the orbital timescale. In practice, this means that the planetesimal size must significantly exceed $s\gg1$ cm \citep{Weidenschilling1977_1}. Modern calculations of planetesimal formation suggest that planetesimals grow very rapidly to radii of order $s\sim100$ km \citep{Johansen2007_1} and do not experience severe interactions with the nebula \citep{Adachi1976_1}. Under such conditions, the entire solid component of the nebula swept up by Jupiter's resonances will be captured and transported inwards. In particular, if one adopts the conservative assumption of a minimum-mass Solar nebula, and an oft-quoted solid-to-gas ratio of $\sim1\%$, the total amount of mass swept up by the resonances is $\sim10 - 20$ Earth masses, about an order of magnitude greater than the cumulative mass of the terrestrial planets.

\begin{figure}[t]
\includegraphics[width=0.95\columnwidth]{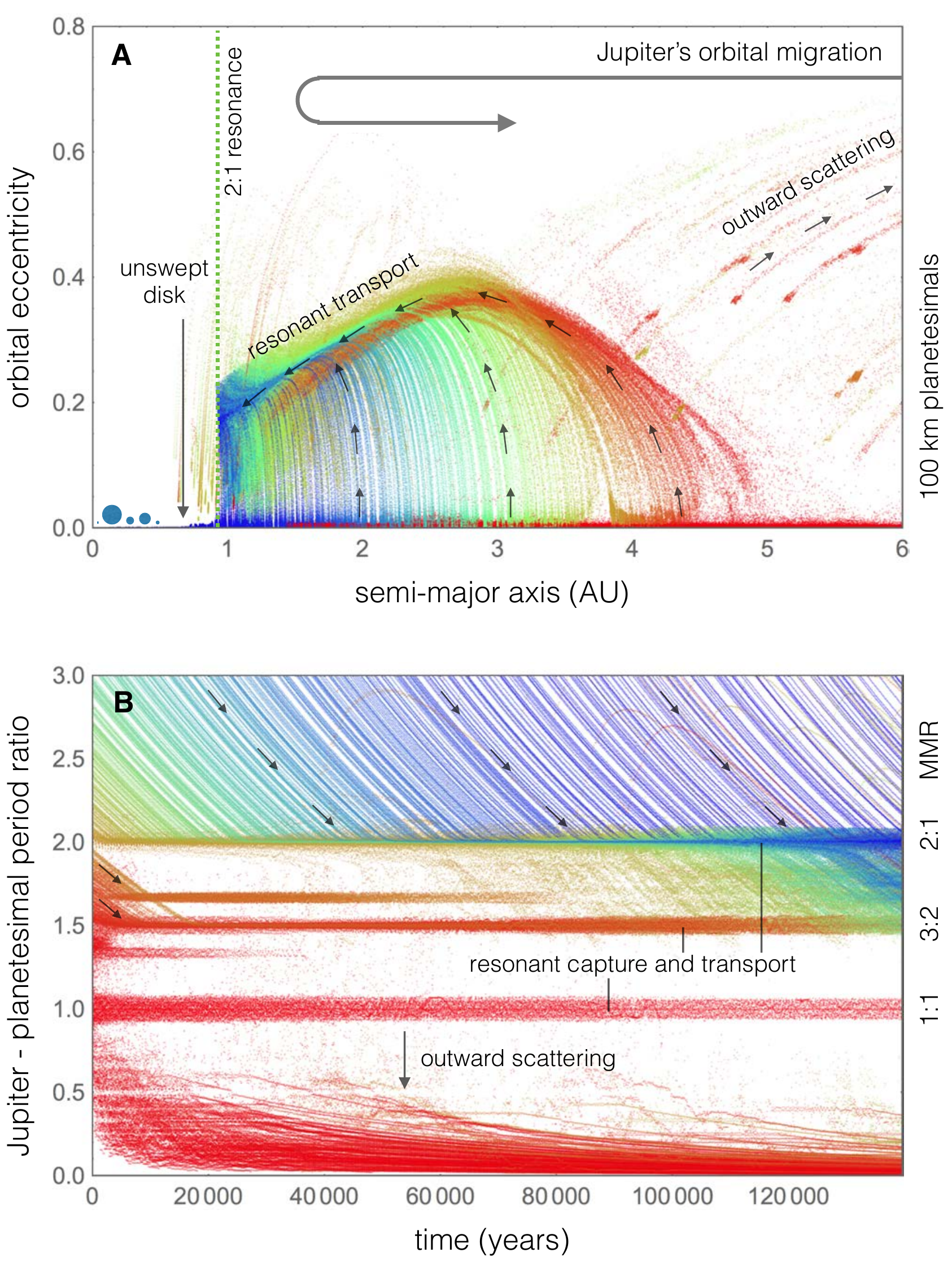}
\caption{\small{Orbital evolution of planetesimals embedded in the Solar nebula, under the effects of a migrating Jupiter. As Jupiter moves inwards from $6$ AU to $1.5$ AU, planetesimals are swept up by mean motion resonances (MMRs). Panel A of this figure shows the increase in the planetesimal eccentricity associated with resonant transport. Note that at the end of the Jupiter's trek, there exists a strong enhancement in the planetesimal density at the Jovian 2:1 MMR. Panel B depicts the preferential population of Jupiter's interior MMRs. Each planetesimal in the simulation is color-coded in accord with its initial condition, and the resultant curves track the orbital excursions of the small bodies as Jupiter's orbit shrinks. Jupiter's return to $\sim5$ AU is not modeled directly. 
In the presented simulation, we assumed a planetesimal size of $s=100$ km. Similar figures corresponding to $s=10$ km and $s=1000$ km can be found in the SI.} }
\label{fig2}
\end{figure}

A planetesimal that is driven inward by resonant migration experiences a concomitant increase in its orbital eccentricity. This effect is generally well understood and stems from adiabatic invariance (see Supporting Information (SI) for a discussion). The eccentricity of a planetesimal embedded in a gaseous nebula cannot grow indefinitely however, as it is damped by aerodynamic drag. Thus, the adiabatic excitation of eccentricity typically stalls at a finite value dependent on the planetesimal size (which controls the magnitude of the dissipative forcing).

We have calculated the orbital evolution of $s=10$ km, $100$ km, and $1000$ km planetesimals as they are swept up by an inward-migrating Jupiter. The results of calculations corresponding to the nominal case of $s=100$ km are presented\footnote{See SI for similar calculations corresponding to $s=10$ km and $1000$ km planetesimals.} in Figure (2). The simulations suggest that for the aforementioned planetesimal sizes, the equilibrium eccentricities are $e\sim0.2,\,0.3$ and $0.5$ respectively. Therefore, substantial orbital crossing will occur between planetesimals caught in resonances with Jupiter and those residing within as-yet unswept regions of the disk. 

The simulations reported in this work were terminated upon Jupiter's arrival to $1.5$ AU. As already mentioned earlier, a resonant encounter with Saturn followed by reversal of migration is envisaged to have occurred subsequently. However, outward migration is not important to the problem at hand because interior material cannot be transported to wider orbits by this process. 

Solid bodies on crossing orbits within densely populated disks experience collisions, which can result in either accretion or fragmentation. The outcome is principally determined by the specific energy of the impact: if this quantity exceeds a critical value characteristic of catastrophic disruption, the target is shattered into two or more pieces \citep{BenzAsphaug1999_1,LeinhardtStewart2009_1}. Adopting parameters appropriate for high-velocity impacts among strong basaltic objects in the gravity-dominated regime (see SI for these parameters), we find analytically that the specific impact energy safely exceeds its threshold value across the range of planetesimal sizes invoked above for impactor-to-target mass ratio of $\sim0.1$ or greater (see SI). In other words, our results suggest that even though one may expect that the real planetesimal disk will harbor a distribution of planetesimal sizes, the trade-off between size-dependent orbital excitation and threshold impact energy leads to an environment where objects of any size above $s\gtrsim 10$ km can be destroyed by bodies that are $\sim 10$ times less massive. Jupiter's resonant shepherding of planetesimals thus initiates a collisional cascade \citep{KesslerCourPalais} that grinds down the planetesimal population to smaller sizes. 

Although the details of resonantly-forced collisional grinding can be complex, an important feature of this process is that once the size of a given planetesimal population is diminished to a point where the effects of aerodynamic drag become important (for example, $s\lesssim1$ km at $1$ AU), the planetesimals will experience a runaway inwards drift \citep{Adachi1976_1}. Importantly, the same process facilitates the removal of material from Jovian resonances and thereby yields a critical planetesimal size below which collisional grinding subsides. Thus, the collisional cascade initiates well before Jupiter reaches its innermost tack, and proceeds as long as Jupiter's migration direction is inward.

Given the exceptionally large impact frequency expected within a mature protoplanetary disk and the dominantly destructive nature of collisions discussed above, we expect that a sizable fraction, if not all, of the transported population of planetesimals will be disrupted and undergo rapid orbital decay following Jupiter's reversal of migration direction. This feature is of critical importance for explaining the Solar System's lack of close-in Super-Earths.  

\section{Decay of Primordial Close-in Planets}
The dominant formation channel (distant formation followed by extensive inward migration \citep{TerquemPapaloizou2007} vs. in-situ conglomeration \citep{ChiangLaughlin2013,HansenMurray2013}) for extrasolar Super-Earths remains controversial. However, a generally agreed-upon framework of core-nucleated accretion of giant planets dictates that the formation of solid multi-Earth-mass cores precedes the formation of giant planets \citep{Pollack1996_1}. Thus, given that the formation of tightly packed close-in systems is ubiquitous in the galaxy, it can be reasonably speculated that at the time of Jupiter's inward journey, a similar population of first-generation planets existed in the Solar System. 
If planets formed from such material, however, they were destroyed.

In exactly the same way as an inward-migrating Jupiter captures planetesimals into resonance, inward-migrating planetesimals will lock into resonance with close-in planets. Provided that the cumulative mass of the resonant planetesimal population is not negligible compared to the mass of the close-in planets, the planetesimals will gravitationally shepherd the close-in planets into the Sun. In other words, the inward-then-outward migration of Jupiter in the early Solar System wiped the inner Solar System's slate clean, setting the stage for the formation of a mass-depleted, gas-starved second generation of terrestrial planets \citep{Morby2012,Chambers2011}. Indeed, within the framework of this picture, the material from which the Earth formed is either the remainder of the violent collisional avalanche, or has been largely emplaced by Jupiter's outward migration. 

\begin{figure}[t]
\includegraphics[width=0.95\columnwidth]{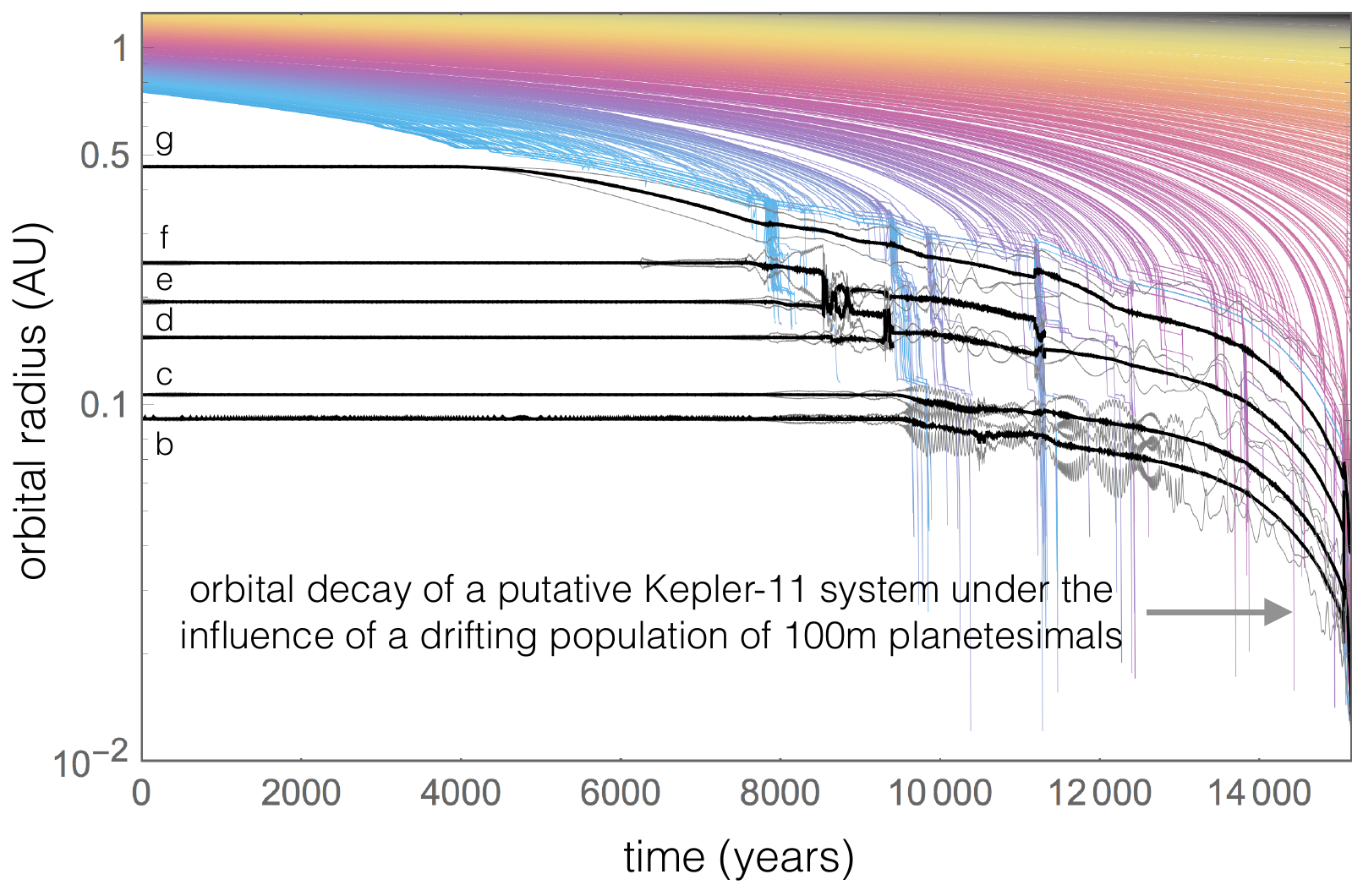}
\caption{\small{Orbital decay of a hypothetical compact system of Super-Earths (an analog of the \textit{Kepler}-11 system) residing within the terrestrial region of the primordial Solar System. Following a collisional avalanche facilitated by Jupiter's migration, a population of planetesimals (here assumed to be ground down to $s=100$ m) decays inwards and resonantly shepherds the interior planets into the star. Planetesimal orbits are shown with colored lines, while the planetary orbits are shown with black and gray lines. Specifically, the planetary semi-major axes are shown in black while the perihelion and aphelion distances are shown in gray. Note that the results shown herein are largely independent of planetesimal size, as long as the planetesimals are small enough to drift inwards on a timescale smaller than $\sim 1$ Myr due to aerodynamic drag.} }
\label{fig3}
\end{figure}

To illustrate the above process, we examined the dynamical evolution of the $Kepler-$11 planetary system\footnote{This example is used for definitiveness. We are not suggesting that a primordial population of the Solar System's close-in planets would have necessarily borne any similarity to the $Kepler-$11 system.} when placed within the inner edge of the Solar nebula, and under the gravitational influence of an extensive population of exterior, inward-drifting planetesimals. The computed evolutionary sequence is shown in Figure (3). Clearly, dissipative resonant transport provides an efficient mechanism for driving close-in planets into the central star. Indeed, the sequence of events associated with Jupiter's so-called ``Grand Tack''  may well have constituted a veritable grand attack on the Solar System's original population of short-period Super-Earths.

\section{Discussion}
This scenario provides a natural explanation for why the inner Solar System bears scant resemblance to the ubiquitous multi-planet systems discovered by the Doppler velocity surveys and by the \textit{Kepler} mission. Moreover, the physical processes that we invoke (namely giant planet migration, collisional disruption of planetesimals, aerodynamic drag, and resonant shepherding) are generic. In consequence, the mechanism described herein should also operate within a non-negligible fraction of extrasolar planetary systems. Accordingly, a series of observational predictions can be formulated. 

First, our calculations imply a strong anti-correlation between the existence of multiple close-in planets and giant planets at orbital periods exceeding $\sim100$ days within the same system. The existing exoplanet catalog is not yet sufficiently detailed to test this hypothesis \citep{Wright2011}. However, direct assessment of the validity of this prediction will be provided by the upcoming TESS and K2 missions. Second, the spectral energy distributions of protoplanetary disks hosting gap-opening planets should exhibit strong infra-red enhancements \citep{Kral2014}, as a consequence of collisional heating and the associated production of dust. Moreover, dust emission morphologies in such disks could in principle exhibit asymmetrical structure \citep{vaderMarel2013}. Most dramatically, our work implies that the majority of Earth-mass planets are strongly enriched in volatile elements and are uninhabitable.


\begin{acknowledgments}
We thank Dave Stevenson, Peter Goldreich, Mike Brown, Geoff Blake, Christopher Spalding, Alessandro Morbidelli and Aurelien Crida for enlightening conversations.
\end{acknowledgments}

\newpage
{\Large Supplementary Information}
\\
\\

\chapter{\textit{Methodology}}

Numerical simulations reported in this work made use of a modified \texttt{mercury6} N-body integration software package \citep{Chambers1999}. The standard gravitational dynamics solver was augmented with the auxilirary effects of aerodynamic drag \citep{Weidenschilling1977,Adachi1976} for small objects as well as fictitious forces that mimic disk-driven migration and circularization of planets. The hybrid symplectic Bulirsch-Stoer algorithm \citep{Press1992} was utilized throughout. \\

\chapter{\textit{Inward Migration of Jupiter}}

Because of dynamical evolution that occurred during the nebular epoch of the Solar System's lifetime, Jupiter's current orbital radius is not informative of its formation site. However, theoretical arguments suggest that the formation of giant planets generally occurs beyond the snow-line i.e. at a distance of $\sim$ a few astronomical units \citep{Stevenson1982, Pollack1996}. Accordingly, in this work we take Jupiter's initial semi-major axis to be $(a_{\rm{J}})_{\rm{i}}=6$ AU.

We employ the standard assumption that the viscosity of the Solar nebula was not overwhelmingly large, meaning that upon formation Jupiter opened a substantial gap in the protoplanetary disk and subsequently migrated inwards in the so-called ``type-II" regime \citep{CridaMorbidelli2007}. Because type-II migration is associated with the global evolution of the disk, the characteristic migration timescale 
\begin{align}
\label{taua}
\frac{2}{\tau_{a}} =\frac{1}{a} \frac{d a}{ dt}
\end{align}
is synonymous with the viscous timescale of the disk. Thus, in agreement with published hydrodynamical simulations, we adopted a value of $(\tau_{a})_{\rm{J}} = 2 \times 10^5$ years \citep{Walsh2011}, corresponding to a Shakrua-Sunayev $\alpha \simeq 0.003$. 

To facilitate migration, we introduced an acceleration $\textbf{a}_{\rm{mig}}$ of the form \citep{PapaloizouLarwood2000}:
\begin{align}
\label{amig}
\mathbf{a}_{\rm{mig}} = - \frac{ \mathbf{v}}{\tau_{a}},
\end{align}
where $\mathbf{v}$ is the orbital velocity. The simulations were terminated when Jupiter reached its envisioned tacking semi-major axis of $(a_{\rm{J}})_{\rm{f}}=1.5$ AU.\\

\chapter{\textit{Planetesimal Evolution}}

As shown in Figure (2) of the main text, a dominant fraction of the planetesimal swarm swept by Jupiter's interior mean-motion commensurabilities was captured into the corresponding resonances. The conditions for resonant capture of small objects initially residing interior to Jupiter's orbit can be deduced analytically within the framework of the circular restricted three body problem \citep{Henrard1982}. Ignoring the effects of Jupiter's minuscule eccentricity, adiabatic theory dictates that capture into a $k:k-1$ resonance is guaranteed if the initial planetesimal eccentricity is less than a critical value:
\begin{align}
\label{ecrit}
e \leqslant \sqrt{6} \left( \frac{3}{\zeta_k} (1-k)^{4/3} k^{2/3} \frac{M_{\odot}}{M_{\rm{J}}} \right)^{-1/3},
\end{align}
where $\zeta_k$ is a interaction coefficient that although different for each resonance, is typically of order unity (for interior 2:1 and 3:2 resonances, $\zeta_2 = -1.19$ and $\zeta_3 = -2.02$ respectively) \citep{MD99}. It can be easily checked that provided reasonable parameter choices, this condition\footnote{Quantitatively, the critical eccentricity evaluates to $e_{\rm{crit}} \simeq 0.15$ and $0.12$ for 2:1 and 3:2 resonances respectively.} is almost certainly satisfied in the primordial Solar nebula. 

An additional requirement for capture to be certain is the fulfillment of the adiabatic condition. For the problem at hand, this means that the orbital convergence timescale must be longer than the resonant libration period, defined as \citep{BatyginMorbidelli2013}:
\begin{align}
\label{omega}
&P_{\rm{lib}} = \frac{2 \pi}{n} \bigg[ \left(\frac{k-1}{k}\right)^{4/3} \left( \frac{M_{\rm{J}}}{M_{\odot}} \right)^2 \frac{\zeta_k^2}{e^2} \nonumber \\
&- 3 (k-1)^2 \left(\frac{k-1}{k} \right)^{2/3} \left( \frac{M_{\rm{J}}}{M_{\odot}} \right) \zeta_k e \bigg]^{-1/2},
\end{align}
where $e$ explicitly refers to the eccentricity at the libration center (i.e. resonant equilibrium).


The first resonant trajectories (associated with the emergence of a homoclinic curve in phase space) appear at an equilibrium eccentricity of
\begin{align}
\label{eres}
&e = \frac{2}{3^{1/3}} \left( \frac{\zeta_k^3}{(\sqrt{k}(k-1))^4} \right)^{1/9} \left(\frac{M_{\rm{J}}}{M_{\odot}} \right)^{1/3}.
\end{align}
The corresponding resonance width is given by
\begin{align}
\label{reswidth}
&\left| \frac{\Delta a}{a} \right| = \frac{8}{3^{2/3}} \left( \frac{(k-1) \, \zeta_k^6}{k^4} \right)^{1/9} \left( \frac{M_{\rm{J}}}{M_{\odot}} \right)^{2/3}.
\end{align}
Accordingly, setting the dissipative resonance crossing time equal to the libration period, we obtain the criterion for adiabatic evolution:
\begin{align}
\label{adcrit}
&\tau_a \gtrsim \frac{2 \pi/ n_{\rm{J}}}{4 \, (3)^{2/3} (\sqrt{k} (k-1))^{2/9} } \left(\frac{\zeta_k M_{\rm{J}}}{M_{\odot}} \right)^{-4/3}.
\end{align}
Quantitatively, the above expression dictates that in order for resonant capture to be ubiquitously successful, the migration timescale must exceed $\tau_a \gtrsim 10^4$ years. Indeed, the typically quoted range of type-II migration timescales is in agreement with the adiabatic condition \citep{Henrard1982}.

\begin{figure}[t]
\includegraphics[width=1\columnwidth]{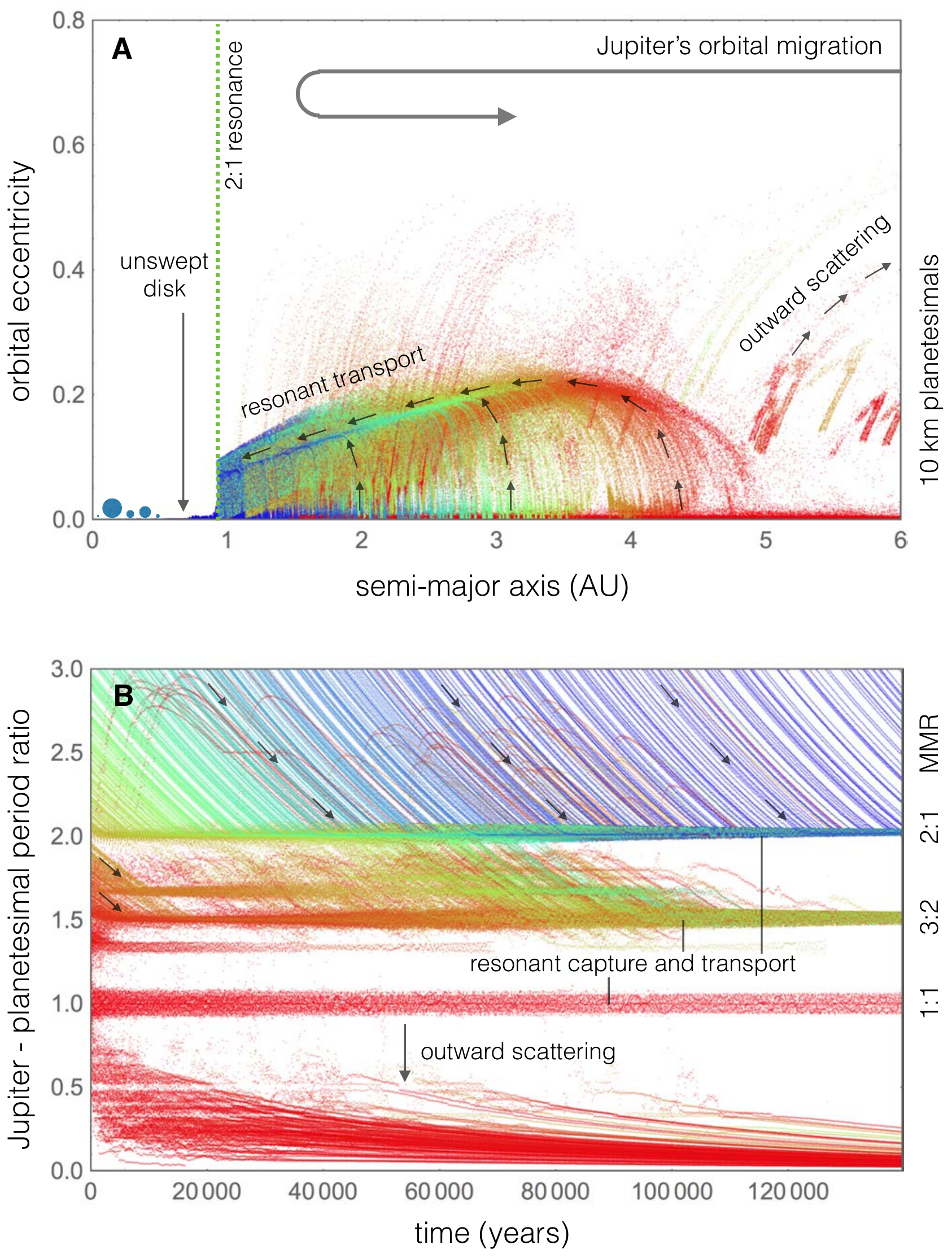}
\caption{\small{Evolutionary sequence of $s=10$ km planetesimals. This figure is an analog of Figure (2) of the main text.}} 
\label{SIFig1}
\end{figure}

Small bodies trapped in resonances will maintain a constant period ratio and will therefore migrate inwards along with Jupiter. Associated with this migration is an adiabatic invariant \citep{YuTrem2001,Peale1986}
\begin{align}
\label{adinv}
\sqrt{a}\left[k - (k-1) \sqrt{1-e^2} \right] = \rm{const.}
\end{align}
The conservation of this quantity implies that as the particle semi-major axis is decreased, its eccentricity must grow. 

By taking a derivative of the above expression, we can obtain a differential equation for test particle's eccentricity growth. Combined with the requirement that the planetesimal semi-major axis must decrease in concert with that of Jupiter [see equation \ref{taua}] once the resonance is established, the differential equation can be solved\footnote{Here, we assume a starting eccentricity of $e=0$.} to obtain the planetesimal eccentricity as a function of time:
\begin{align}
e = \sqrt{\frac{1-\exp(t/\tau_{a})-2k(1 - \exp(t/(2\tau_{a}))}{(k-1)^2}}
\end{align}
This expression implies that the characteristic timescale for any given particle to reach its equilibrium eccentricity is of order $\sim0.1-0.5 \, \tau_{a}$.

Within the context of our numerical simulations, the solid component of the nebula was initialized as a disk comprised of 1000 objects with inner and outer radii of $(a_{\rm{disk}})_{\rm{in}}=0.1$ AU and $(a_{\rm{disk}})_{\rm{out}}=6$ AU respectively. The outer radius of this disk is not indicative of the true outer radius of the Solar nebula. Rather, planetesimals with semi-major axes greater than Jupiter's initial semi-major axis were ignored in our simulations because they cannot be transported inwards by sweeping exterior mean motion resonances (divergent resonant encounters cannot lead to capture) \citep{Henrard1982,MD99}. 

\begin{figure}[t]
\includegraphics[width=1\columnwidth]{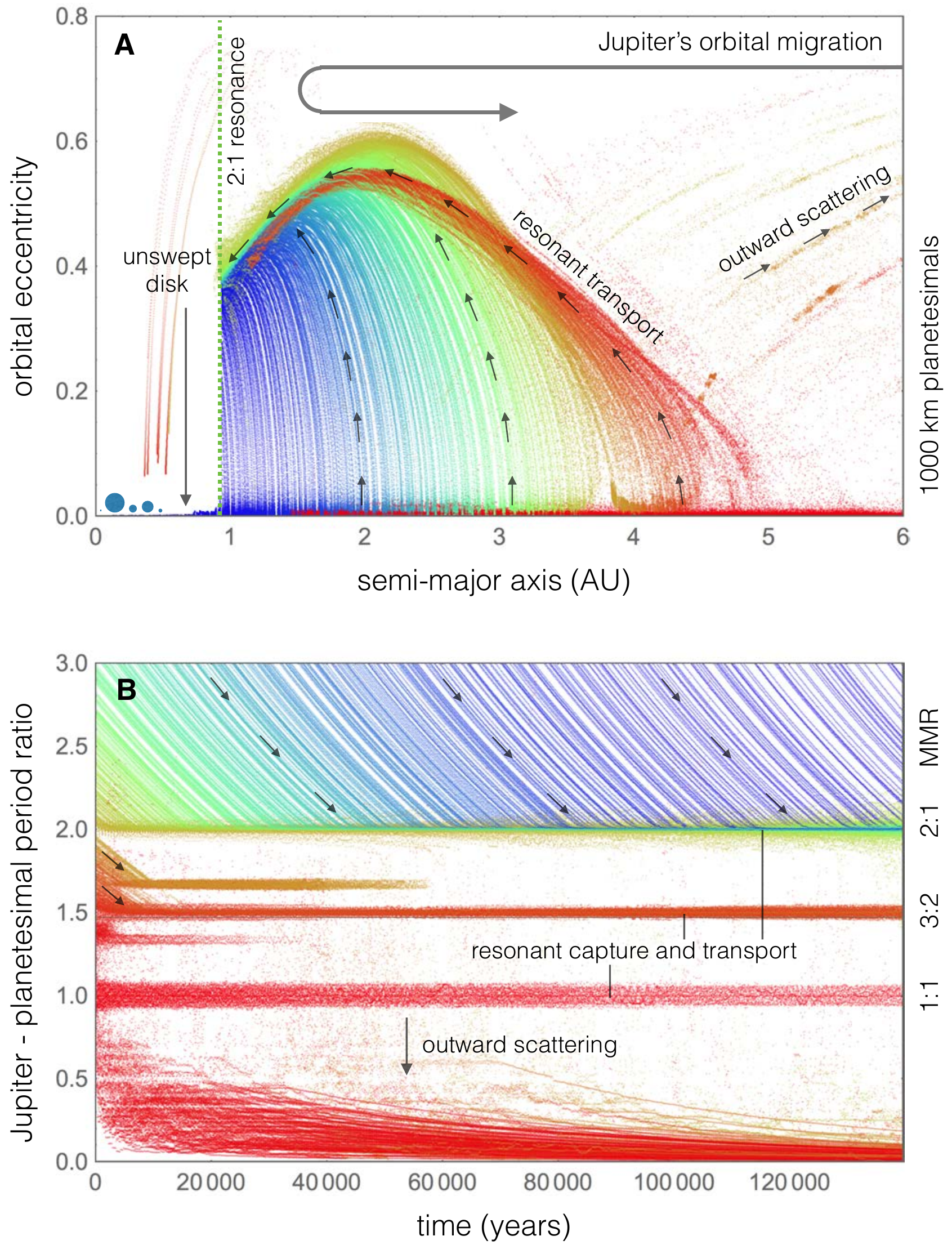}
\caption{\small{Evolutionary sequence of $s=1000$ km planetesimals. This figure is an analog of Figure (2) of the main text.}} 
\label{SIFig2}
\end{figure}

The radial distribution of the planetesimals followed a surface density profile equivalent to that of a Mestel disk \citep{Mestel1963}: 
\begin{align}
\label{surfden}
\Sigma = \Sigma_0 \left( \frac{a}{a_0} \right)^{-1},
\end{align}
where $\Sigma_0$ is the disk surface density at a semi-major axis $a_0$. We note that the total planetesimal mass transported inwards by resonant shepherding (assuming that first-order resonances such as 2:1 and 3:2 are primarily responsible for long-range migration) is
\begin{align}
M_{\rm{tot}} &= 2 \pi  \, \epsilon \, f  \int^{(a_{\rm{J}})_{\rm{i}}/2^{2/3}}_{(a_{\rm{J}})_{\rm{f}}} \Sigma_0 \left( \frac{a_0}{a} \right) a \, da \nonumber \\
&\simeq 2 \pi  \, \epsilon \, f \, \Sigma_0 \, a_0 \, (a_{\rm{J}})^{\rm{i}}/2^{2/3},
\end{align}
where $f \sim 0.01$ is the disk solid-to-gas mass ratio and $\epsilon \lesssim 1$ is the resonant transport efficiency\footnote{Note that $\epsilon$ is primarily limited by chaotic diffusion and the associated removal of objects from Jovian resonances \citep{Wisdom1980}.}. Quantitatively, only $\sim20\%$ of the captured objects were observed to chaotically escape the resonances during the entirety of Jupiter's inward trek in the simulations. 

Adopting MMSN-like parameters of $\Sigma_0 \simeq 2000$ g/cm$^2$ at $a_0 = 1$AU, we obtain $M_{\rm{tot}} \sim 20 M_{\oplus} \ll M_{\rm{J}}$. Consequently, the gravitational back-reaction of the planetesimals onto an inward-migrating Jupiter is almost certainly unimportant, and for the purposes of this set of simulations we ignored this effect. Correspondingly, Jupiter's eccentricity remained at a near-null value throughout the integrations.

Assuming that the solid and gaseous component of the nebula follow the same surface density profile [\ref{surfden}], radial pressure support will yield a sub-Keplerian circular velocity profile of the form \citep{Weidenschilling1977}:
\begin{align}
\mathbf{v}_{\rm{gas}} = v_{\rm{K}} \sqrt{1 - 3\frac{c_{\rm{s}}^2}{v_{\rm{K}}^2}} \, \hat{\varphi} = v_{\rm{K}} (1 - \eta) \, \hat{\varphi},
\end{align}
where $c_{\rm{s}}$ and $v_{\rm{K}}$ are the sound and Keplerian speeds respectively, while $\hat{\varphi}$ denotes the azimuthal unit vector. Prototypical disk parameters yield a value of $\eta$ in the range $0.001 - 0.01$. Accordingly, in this work we adopted $\eta=0.005$ (corresponding to a disk aspect ratio of $h/r \simeq 0.05$) for all calculations.

Planetesimals embedded in a sub-Keplerian gas disk will experience orbital decay due to aerodynamic drag. The hydrodynamic Reynolds number relevant for planetesimals bigger than $s \gtrsim 10$ m substantially exceeds unity \citep{Malhotra1993}. In this regime, drag acceleration $\mathbf{a}_{\rm{drag}}$ is quadratic in velocity and independent of fluid viscosity \citep{LandauLifshitz1959}:
\begin{align}
\label{adrag}
\mathbf{a}_{\rm{drag}} = - \frac{\pi \, \mathcal{C}_{\rm{D}}}{2 m} s^2 \rho_{\rm{gas}} v_{\rm{rel}} \mathbf{v}_{\rm{rel}}.
\end{align}
In the above expression, $\mathbf{v}_{\rm{rel}} = \mathbf{v} - \mathbf{v}_{\rm{gas}}$ is the relative velocity between the planetesimal and the gas, whereas $m$ is the planetesimal mass. For the entire set of our simulations, the aforementioned acceleration was implemented with a drag coefficient of $\mathcal{C}_{\rm{D}} \simeq 0.5$. 

In addition to the decrease in semi-major axes, the introduction of this acceleration results in circularization of the orbits and damping of the vertical motion (inclinations). To leading order in $e$ and $i$, the rates of these dissipative effects are expressed as follows \citep{Adachi1976}:
\begin{align}
\label{dadtdedt}
\frac{1}{a}\frac{da}{dt} &= -\frac{2}{\tau_0} \eta \sqrt{\frac{5}{8} e^2 + \frac{1}{2} i^2 + \eta^2 } \nonumber \\
\frac{1}{e}\frac{de}{dt} &= \frac{2}{i}\frac{di}{dt} = -\frac{1}{\tau_0} \sqrt{\frac{5}{8} e^2 + \frac{1}{2} i^2 + \eta^2 },
\end{align}
where
\begin{align}
\label{tau0}
\tau_0 = \left( \frac{\pi \, \mathcal{C}_{\rm{D}}}{2 m} s^2 \rho_{\rm{gas}} v_{\rm{K}} \right)^{-1}
\end{align}

In light of equations [\ref{dadtdedt}], it is important to recall that although first order resonant interactions with Jupiter will modulate the planetesimal eccentricity [equation \ref{adinv}], they will not excite orbital inclination \citep{MD99}. Aerodynamic drag on the other hand, tends to damp vertical motion of eccentric orbits. Thus, the process of resonant transport leads to a dissipative confinement of the solid component of the nebula to the mid-plane of the disk. 

Note that here, we are assuming that turbulent fluctuations in the disk are not sufficiently large to obstruct resonant capture. This is likely a safe assumption given that the orbital plane of planetesimals will likely be confined to the dominantly laminar mid-plane. However, an improved iteration of the considered model could in principle account for this effect directly. 

A promising theory for the dominant channel for planetesimal formation invokes the gravitational collapse of solid over-densities within the disk \citep{Johansen2007}, which are in turn fabricated by turbulent forcing and streaming instabilities \citep{YoudinGoodman2005}. Within the framework of this picture, planetesimals are born big, with peak masses of order that of Ceres. As a typical planetesimal size, it is sensible to choose $s = 100$ km and simulations employing this default value are reported in the main text. For completeness, we have repeated the numerical experiments of inward migration of Jupiter and the associated transport of planetesimals with $s = 10$ km and $s = 1000$ km. The corresponding results are shown in Figures (4) and (5) respectively. 

As expected, hydrodynamical damping of smaller objects is substantially more pronounced, which means that resonant transport is more efficient at eccentricity excitation for higher-mass bodies. Accordingly, our simulations show that the equilibrium eccentricity attained by $10$ km and $1000$ km objects throughout their inward trek is $e \simeq 0.2$ and $e \simeq 0.5$ respectively. \\

\chapter{\textit{Collisions}}

Resonant shepherding of planetesimals by Jupiter will lead to a concentration of small objects at particular (mean-motion resonant) orbits. In turn, such confinement will lead to a strong enhancement in the rate of collisions among planetesimals. A rough estimate of the collision frequency, $\nu$ (defined as the inverse of the mean time between collisions experienced by a single particle), can be obtained from a simple $n\, \sigma \, v$ calculation. Specifically, for the problem at hand we have \citep{Armitage2010}: 
\begin{align}
\label{nsigmav}
\nu = \frac{M_{\rm{tot}}/m}{2 \pi \langle e \rangle \tan\langle i \rangle a^3}\, \pi s^2 v_{\rm{K}} \langle e \rangle.
\end{align}

If we take the entire planetesimal population to be comprised of $s = 100$ km planetesimals, adopt $\langle i \rangle \sim 10^{-4}$ as suggested by the simulations\footnote{Note that the smallest physically sensible value of the inclination that we can adopt for this calculation is $i_{\rm{min}} = \tan^{-1}(s/a) \sim 10^{-6}$.}, and $M_{\rm{tot}} \sim 10M_{\oplus}$, at $a = 1$ AU we obtain $\nu \sim 0.05$. This means that each member of the planetesimal population can be expected to suffer a collision each 20 orbits or so. This constitutes an exceptionally large collision rate and allows one to infer that collisional grinding of the small bodies initiated by Jupiter will be efficient. 

The detailed description of the outcome of collisions can in general be complex and may depend on numerous parameters that are specific to a particular impact. In an averaged sense, however, the controlling factor in determining the result of collisions is the specific energy \citep{Armitage2010}:
\begin{align}
\label{Q}
Q=\left( \frac{m'}{M} \right) \left(\frac{v_{\rm{enc}}^2}{2} \right).
\end{align}
In this equation, $m'$ is the impactor mass, $M \geqslant m'$ is the target mass, and $v_{\rm{enc}}$ is the encounter velocity. 

There exists a critical value $Q_{\rm{D}}^*$ characteristic of catastrophic disruption, such that if $Q$ exceeds $Q_{\rm{D}}^*$, the target is shattered into two or more pieces of mass no greater than $M/2$. In the gravity-dominated regime (appropriate for bodies with radius $s \gtrsim 0.5$ km), the expression for $Q_{\rm{D}}^*$ reads \citep{BenzAsphaug1999,LeinhardtStewart2009}:
\begin{align}
Q_{\rm{D}}^*= q\rho\left(\frac{ R}{1\, \rm{cm}} \right)^{b}.
\end{align}

Adopting $\rho=3$ g/cm$^3$, $q=0.5$ and $b=1.36$ appropriate for high-velocity impacts among basaltic objects and $v_{\rm{enc}} \sim e\,v_{\rm{kep}}$ for the encounter velocity, we find that at 1AU $Q \gtrsim Q_{\rm{D}}^*$ for mass ratios of $(m'/M) \gtrsim 0.01, 0.03$ and $0.1$ for $s = 10, 100$ and $1000$ km respectively. Given that the process of collisional fragmentation acts to produce more objects of diminished sizes, the above estimates suggest that the cascade enabled by Jupiter's inward migration will efficiently grind down the planetesimal population to sizes where orbital decay due to aerodynamic drag becomes an efficient removal mechanism. 

It should be noted that the simple calculations shown above ignore a number of detailed effects that can be taken into account within the framework of a more sophisticated model. For example, we have ignored the various geometrical particularities of collisions within resonant orbits, the associated collisional damping and viscous stirring, as well as the time-evolution of the planetesimal size distribution and the self-consistent removal of collisional fragments from Jovian resonances by rapid inward drift. While these issues may be of substantial interest as an avenue for follow-up work, they are not central to the arguments presented here, and may therefore be omitted for the purposes of this study.\\

\chapter{\textit{Orbital Decay of a Kepler-11 Analog}}


\begin{figure}[t]
\includegraphics[width=1\columnwidth]{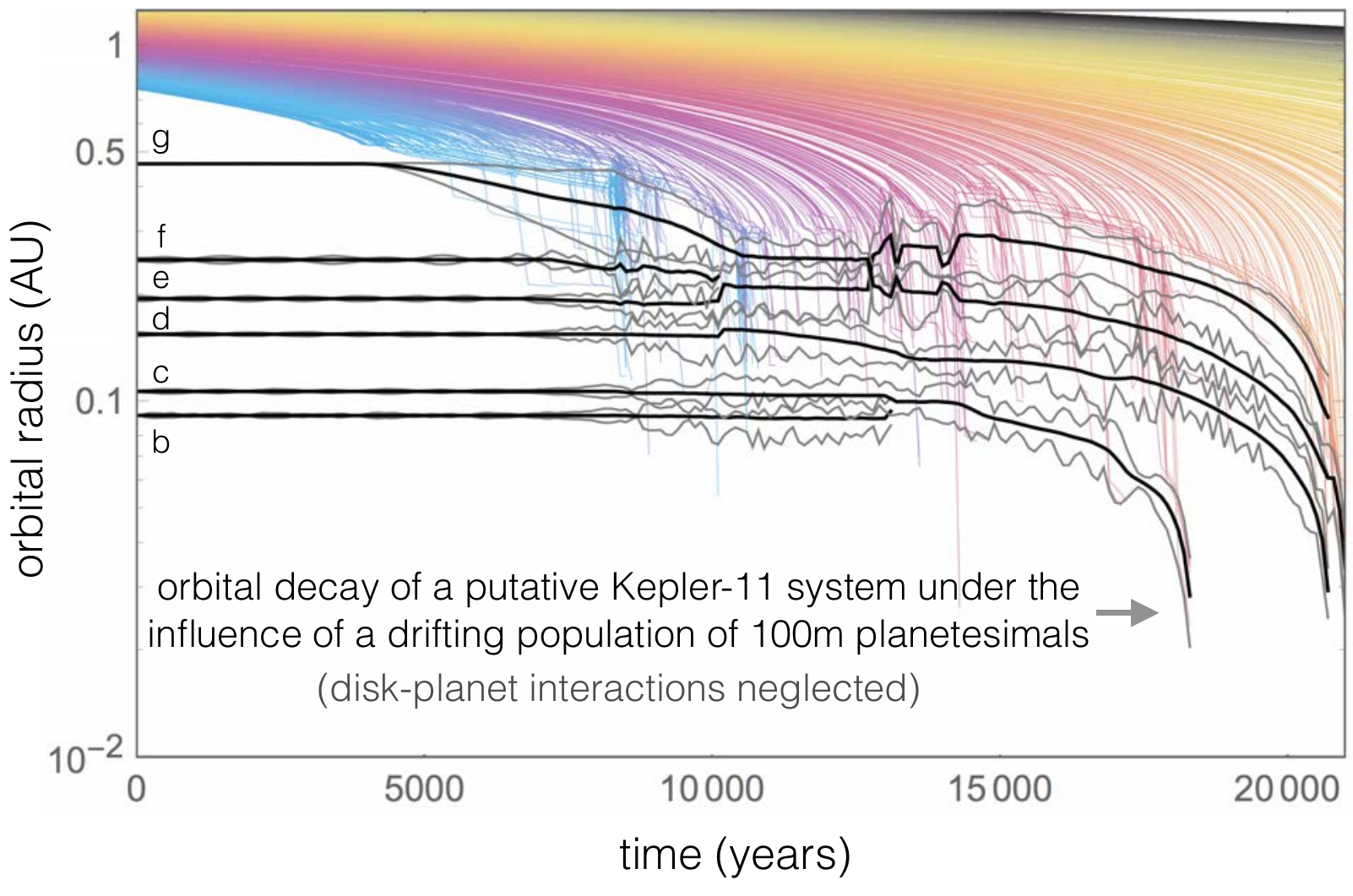}
\caption{\small{Orbital decay of a clone of the \textit{Kepler}-11 system embedded within the inner Solar nebula. The calculation presented herein mirrors that shown in Figure (3) of the main text. However, within the framework of this simulation, dissipative interactions between the planets (but not planetesimals) and the disk are entirely neglected. Evidently, the effect of eccentricity damping on the planets plays a negligible role in determining the qualitative behavior of the system.}} 
\label{SIFig3}
\end{figure}

Extended inward flow of collisionally disrupted planetesimals may be halted (by resonant capture) if the migration route is obstructed by at least one massive planet \citep{Malhotra1993}. However, if the cumulative mass of planetesimals trapped in exterior planetary resonances becomes substantial (i.e. non-negligible with respect to the planetary mass), inward migration of the entire system will occur. In light of this notion, we performed numerical experiments aimed at evaluating the long-term fate of a compact, close-in aggregate of sub-Jovian planets embedded into the terrestrial region of the Solar System, under the influence of an exterior flood of inward-spiraling planetesimals.

Within the context of these simulations, the dynamical evolution of small bodies followed the same framework as that discussed above. The (presumed collisionally evolved) swarm of planetesimals was initialized as a circular annulus of 1000 objects spanning $0.75-1.25$ AU, and allowed to drift inwards. The total mass of the swarm was taken to be $M_{\rm{tot}} = 20 M_{\oplus}$. 

For the purposes of computing the effects of aerodynamic drag, a characteristic radius of $s = 0.1$ km was assumed for the entire population. We note that this value sets the timescale on which orbital decay takes place. In other words, similar dynamical evolution can be obtained by choosing a radius of $s = 1$ km and running the integrations for $10$ times longer. 

As before, the maximal (aerodynamically-forced) orbital decay rates that satisfy the adiabatic condition can be calculated using classical perturbation theory \citep{MD99}. Assuming that the planetary orbit is circular, the libration period of a $k:k-1$ exterior resonance is \citep{BatyginMorbidelli2013} 
\begin{align}
\label{omegaext}
P_{\rm{lib}} &= \frac{2\pi}{n} \bigg[ \left( \frac{\tilde{m}}{M_{\oplus}} \right)^2 \frac{\xi_k^2}{e^2} - 3 k^2 \left( \frac{\tilde{m}}{M_{\oplus}} \right) \xi_k e \bigg]^{-1/2},
\end{align}
where $\tilde{m}$ is the planetary mass, and (akin to $\zeta_k$) $\xi_k$ is a coefficient of order unity (for exterior 2:1 and 3:2 resonances,  $\xi_2 = 0.428$ and $\xi_3 = 0.515$ respectively). 


The equilibrium eccentricity at the inception of the resonance is
\begin{align}
\label{eeqext}
&e = \frac{2}{3^{1/3}} \left( \frac{\xi_k \, \tilde{m}}{ k^2 \, M_{\odot}} \right)^{1/3},
\end{align}
with a corresponding resonance width of
\begin{align}
\label{reswidthext}
\left| \frac{\Delta a}{a} \right| = \frac{8}{3^{2/3} k^{1/3}} \left(\frac{ \xi_k \tilde{m} }{M_{\odot}} \right)^{2/3}.
\end{align}
Calculating the resonance crossing time using equation [\ref{dadtdedt}], we obtain the following expression for the adiabatic threshold:
\begin{align}
\label{adcritext}
&\frac{3 \, \mathcal{C}_{\rm{D}}}{8} \left(\frac{\rho_{\rm{gas}}}{\rho} \right) \left( \frac{v_{\rm{K}} \, \eta^2}{s} \right) \lesssim n \bigg( \frac{3^{2/3} (k-1)}{\pi \, k^{2/3}} \bigg) \left( \frac{\xi_k \tilde{m}}{M_{\odot}} \right)^{4/3}.
\end{align}

As an example, equation [\ref{adcritext}] dictates that adiabatic capture into a first order resonance with a $\tilde{m}=10M_{\oplus}$ planet residing in a MMSN at $a=0.4$ AU is assured for particles with radii in excess of $s \gtrsim 1$ km. While this is a generous constraint already, we note further that even in a trans-adiabatic regime, resonant capture will still occur, but with a diminished probability \citep{Malhotra1993}. To this end, our numerical simulations intentionally break the adiabatic limit and thereby demonstrate that inward migration of close-in planets can still be successfully forced by debris substantially smaller than $s \lesssim 1$ km.

In direct analogy with the capture of planetesimals into interior Jovian resonances [equation \ref{ecrit}], capture into exterior planetary resonances is assured if the particle eccentricity does not exceed a critical value \citep{MD99}
\begin{align}
e \leqslant \sqrt{6} \left( \frac{3}{\xi_k} k^{2} \frac{M_{\odot}}{\tilde{m}} \right)^{-1/3}.
\end{align}
Given that the hydrodynamic eccentricity damping timescale is $\sim 100$ times shorter than the semi-major axis decay timescale [see equations \ref{dadtdedt}], and the latter process is necessary for resonant capture in the first place, it is sensible to assume that this condition is well satisfied in the inner Solar nebula.

As an emblematic example of a tightly packed set of close-in planets, we adopted the \textit{Kepler}-11 system \citep{Lissauer2011} and inserted it interior to the decaying disk of planetesimals. The planets were initialized on their current near-circular orbits (specifically, we adopted the ``all-eccentric" orbital fit delineated in ref. \citep{Lissauer2011}). Gravitational interactions between the planets and the planetesimals were computed in the conventional N-body fashion. However, the self-gravity of the planetesimal disk was ignored to save computational costs. 

Unlike Jupiter, no fictitious migration due to the disk was imposed on the planets. Damping of the orbital eccentricities due to angular momentum exchange with the disk was implemented via an acceleration of the form \citep{PapaloizouLarwood2000}:
\begin{align}
\label{adamp}
\mathbf{a}_{\rm{damp}} = - 2 \frac{ (\mathbf{v} \cdot \mathbf{r} )}{r^2\tau_{e}} \mathbf{r},
\end{align}
where as in equation [\ref{taua}],
\begin{align}
\label{adamp}
\tau_{e} =\frac{1}{e} \frac{d e}{ dt}.
\end{align}

The quantity $\tau_e$ depends on numerous physical properties of the disk and is in general somewhat uncertain \citep{Ward1997}. In this work, we performed simulations that adopted $\tau_e = 10^3$ years (shown in Figure 3 of the main text) as well as a run with $\tau_e = \infty$ (shown as Figure 6). Examination of the presented orbital solutions show that in either case, only mild eccentricities are attained as the planets are ushered towards the Sun. 

The exact extent of orbital excitation is set by the delicate interplay between adiabatic and dissipative dynamics\footnote{We note that for the non-restricted (i.e. planetary) three-body problem, an adiabatic invariant analogous to equation [\ref{adinv}] can be defined \citep{BatyginMorbidelli2013}.}. However, our results broadly suggest that for the problem at hand, the exact value of $\tau_e$ does not factor into the final answer appreciably. That is, our simulations show that irrespective of the details of the chaotic dynamical evolution that occurs as planetesimals lock into exterior resonance with the planets, the planetary orbits decay onto the host star over a timescale that is short compared with the typical lifetimes of protoplanetary disks. \\

\end{article}


\begin{thebibliography}{10}

\bibitem[Morbidelli et al.(2012)]{Morby2012} Morbidelli, A., Lunine, J.~I., O'Brien, D.~P., Raymond, S.~N., Walsh, K.~J.\ Building Terrestrial Planets.\ Annual Review of Earth and Planetary Sciences 40, 251-275 (2012)

\bibitem[Mayor et al.(2011)]{Mayor2011} Mayor, M., and 13 colleagues \ The HARPS search for southern extra-solar planets XXXIV. Occurrence, mass distribution and orbital properties of super-Earths and Neptune-mass planets.\ ArXiv e-prints arXiv:1109.2497 (2011)

\bibitem[Batalha et al.(2013)]{Batalha2013} Batalha, N.~M., and 75 colleagues \ Planetary Candidates Observed by Kepler. III. Analysis of the First 16 Months of Data.\ The Astrophysical Journal Supplement Series 204, 24 (2013)

\bibitem[Lissauer et al.(2011)]{Lissauer2011_1} Lissauer, J.~J., and 38 colleagues \ A closely packed system of low-mass, low-density planets transiting Kepler-11.\ Nature 470, 53-58 (2011)

\bibitem[Connelly et al.(2012)]{Connelly2012} Connelly, J.~N., Bizzarro, M., Krot, A.~N., Nordlund, {\AA}., Wielandt, D., Ivanova, M.~A.\ The Absolute Chronology and Thermal Processing of Solids in the Solar Protoplanetary Disk.\ Science 338, 651 (2012)

\bibitem[Chambers(2011)]{Chambers2011} Chambers, J.\ Terrestrial Planet Formation.\ Exoplanets, edited by S.~Seager.~ Tucson, AZ: University of Arizona Press, p.297-317 (2011)

\bibitem[Fischer and Ciesla(2014)]{FischerCiesla2014} Fischer, R.~A., Ciesla, F.~J.\ 2014.\ Dynamics of the terrestrial planets from a large number of N-body simulations.\ Earth and Planetary Science Letters 392, 28-38. 

\bibitem[Izidoro et al.(2014)]{Izidoro2014} Izidoro, A., Haghighipour, N., Winter, O.~C., Tsuchida, M.\ 2014.\ Terrestrial Planet Formation in a Protoplanetary Disk with a Local Mass Depletion: A Successful Scenario for the Formation of Mars.\ The Astrophysical Journal 782, 31. 

\bibitem[Hansen(2009)]{Hansen2009} Hansen, B.~M.~S.\ Formation of the Terrestrial Planets from a Narrow Annulus.\ The Astrophysical Journal 703, 1131-1140 (2009)

\bibitem[Hayashi(1981)]{Hayashi1981} Hayashi, C.\ Structure of the Solar Nebula, Growth and Decay of Magnetic Fields and Effects of Magnetic and Turbulent Viscosities on the Nebula.\ Progress of Theoretical Physics Supplement 70, 35-53 (1981)

\bibitem[Desch(2007)]{Desch2007} Desch, S.~J.\ Mass Distribution and Planet Formation in the Solar Nebula.\ The Astrophysical Journal 671, 878-893 (2007)

\bibitem[Boley et al.(2014)]{Boley2014} Boley, A.~C., Morris, M.~A., Ford, E.~B.\ 2014.\ Overcoming the Meter Barrier and the Formation of Systems with Tightly Packed Inner Planets (STIPs).\ The Astrophysical Journal 792, LL27. 

\bibitem[Walsh et al.(2011)]{Walsh2011_1} Walsh, K.~J., Morbidelli, A., Raymond, S.~N., O'Brien, D.~P., Mandell, A.~M.\ A low mass for Mars from Jupiter's early gas-driven migration.\ Nature 475, 206-209 (2011)

\bibitem[Kley and Nelson(2012)]{KleyNelson2012} Kley, W., Nelson, R.~P.\ Planet-Disk Interaction and Orbital Evolution.\ Annual Review of Astronomy and Astrophysics 50, 211-249 (2012)

\bibitem[Masset and Snellgrove(2001)]{MassetSnellgrove2001} Masset, F., Snellgrove, M.\ Reversing type II migration: resonance trapping of a lighter giant protoplanet.\ Monthly Notices of the Royal Astronomical Society 320, L55-L59 (2001)

\bibitem[Morbidelli(2013)]{Morbidelli2013} Morbidelli, A.\ 2013.\ Dynamical Evolution of Planetary Systems.\ Planets, Stars and Stellar Systems.~Volume 3: Solar and Stellar Planetary Systems 63. 

\bibitem[Wright et al.(2011)]{Wright2011} Wright, J.~T., and 10 colleagues\ The Exoplanet Orbit Database.\ Publications of the Astronomical Society of the Pacific 123, 412-422 (2011)

\bibitem[Morbidelli and Crida(2007)]{MorbidelliCrida2007} Morbidelli, A., Crida, A.\ The dynamics of Jupiter and Saturn in the gaseous protoplanetary disk.\ Icarus 191, 158-171 (2007)

\bibitem[Lee and Peale(2002)]{LeePeale2002} Lee, M.~H., Peale, S.~J.\ 2002.\ Dynamics and Origin of the 2:1 Orbital Resonances of the GJ 876 Planets.\ The Astrophysical Journal 567, 596-609. 

\bibitem[O'Brien et al.(2014)]{OBrien2014} O'Brien, D.~P., Walsh, K.~J., Morbidelli, A., Raymond, S.~N., Mandell, A.~M.\ Water delivery and giant impacts in the `Grand Tack' scenario.\ Icarus 239, 74-84 (2014)

\bibitem[Morbidelli et al.(2007)]{Morby2007} Morbidelli, A., Tsiganis, K., Crida, A., Levison, H.~F., Gomes, R.\ Dynamics of the Giant Planets of the Solar System in the Gaseous Protoplanetary Disk and Their Relationship to the Current Orbital Architecture.\ The Astronomical Journal 134, 1790-1798 (2007)

\bibitem[Batygin and Brown(2010)]{BatyginBrown2010} Batygin, K., Brown, M.~E.\ Early Dynamical Evolution of the Solar System: Pinning Down the Initial Conditions of the Nice Model.\ The Astrophysical Journal 716, 1323-1331 (2010)

\bibitem[Malhotra(1993)]{Malhotra1993_1} Malhotra, R.\ Orbital resonances in the solar nebula - Strengths and weaknesses.\ Icarus 106, 264 (1993)

\bibitem[Yu and Tremaine(2001)]{YuTrem2001_1} Yu, Q., Tremaine, S.\ Resonant Capture by Inward-migrating Planets.\ The Astronomical Journal 121, 1736-1740 (2001)

\bibitem[Weidenschilling(1977)]{Weidenschilling1977_1} Weidenschilling, S.~J.\ Aerodynamics of solid bodies in the solar nebula.\ Monthly Notices of the Royal Astronomical Society 180, 57-70 (1977)

\bibitem[Johansen et al.(2007)]{Johansen2007_1} Johansen, A., Oishi, J.~S., Mac Low, M.-M., Klahr, H., Henning, T., Youdin, A.\ Rapid planetesimal formation in turbulent circumstellar disks.\ Nature 448, 1022-1025 (2007)

\bibitem[Adachi et al.(1976)]{Adachi1976_1} Adachi, I., Hayashi, C., Nakazawa, K.\ The gas drag effect on the elliptical motion of a solid body in the primordial solar nebula..\ Progress of Theoretical Physics 56, 1756-1771 (1976)

\bibitem[Benz and Asphaug(1999)]{BenzAsphaug1999_1} Benz, W., Asphaug, E.\ Catastrophic Disruptions Revisited.\ Icarus 142, 5-20 (1999)

\bibitem[Leinhardt and Stewart(2009)]{LeinhardtStewart2009_1} Leinhardt, Z.~M., Stewart, S.~T.\ Full numerical simulations of catastrophic small body collisions.\ Icarus 199, 542-559 (2009)

\bibitem[Kessler and Cour-Palais(1978)]{KesslerCourPalais} Kessler, D.~J., Cour-Palais, B.~G.\ Collision frequency of artificial satellites: The creation of a debris belt.\ Journal of Geophysical Research 83, 2637-2646 (1978)

\bibitem[Terquem and Papaloizou(2007)]{TerquemPapaloizou2007} Terquem, C., Papaloizou, J.~C.~B.\ Migration and the Formation of Systems of Hot Super-Earths and Neptunes.\ The Astrophysical Journal 654, 1110-1120 (2007)

\bibitem[Chiang and Laughlin(2013)]{ChiangLaughlin2013} Chiang, E., Laughlin, G.\ The minimum-mass extrasolar nebula: in situ formation of close-in super-Earths.\ Monthly Notices of the Royal Astronomical Society 431, 3444-3455 (2013)

\bibitem[Hansen and Murray(2013)]{HansenMurray2013} Hansen, B.~M.~S., Murray, N.\ Testing in Situ Assembly with the Kepler Planet Candidate Sample.\ The Astrophysical Journal 775, 53 (2013)

\bibitem[Pollack et al.(1996)]{Pollack1996_1} Pollack, J.~B., Hubickyj, O., Bodenheimer, P., Lissauer, J.~J., Podolak, M., Greenzweig, Y.\ Formation of the Giant Planets by Concurrent Accretion of Solids and Gas.\ Icarus 124, 62-85 (1996)

\bibitem[Kral et al.(2014)]{Kral2014} Kral, Q., Thebault, P., Augereau, J.-C., Boccaletti, A., Charnoz, S.\ Signatures of massive collisions in debris discs. \ Astronomy and Astrophysics, in press (2014)

\bibitem[van der Marel et al.(2013)]{vaderMarel2013} van der Marel, N., and 11 colleagues \  A Major Asymmetric Dust Trap in a Transition Disk.\ Science 340, 1199-1202 (2013)

\end{thebibliography}

\begin{thebibliography}{10}

\bibitem[Chambers(1999)]{Chambers1999} Chambers, J.~E.,\ A hybrid symplectic integrator that permits close encounters between massive bodies.\  \textit{Mon. Not. R. Astron. Soc.} 304, 793-799 (1999)

\bibitem[Weidenschilling(1977)]{Weidenschilling1977} Weidenschilling, S.~J.\ Aerodynamics of solid bodies in the solar nebula. \textit{Mon. Not. R. Astron. Soc.} 180, 57-70 (1977)

\bibitem[Adachi et al.(1976)]{Adachi1976} Adachi, I., Hayashi, C., Nakazawa, K.\ The gas drag effect on the elliptical motion of a solid body in the primordial solar nebula.\ \textit{Progress of Theoretical Physics} 56, 1756-1771 (1976)

\bibitem[Press et al.(2007)]{Press1992} Press W.~H. et al., \textit{Numerical recipes in FORTRAN. The art of scientific computing.} Cambridge: University Press, 2nd ed.\ (1992)

\bibitem[Pollack et al.(1996)]{Pollack1996} Pollack, J.~B., Hubickyj, O., Bodenheimer, P., Lissauer, J.~J., Podolak, M., Greenzweig, Y.\ Formation of the Giant Planets by Concurrent Accretion of Solids and Gas.\ \textit{Icarus} 124, 62-85 (1996)

\bibitem[Stevenson(1982)]{Stevenson1982} Stevenson, D.~J., Formation of the giant planets.\ \textit{Planet. Spac. Sci.} 30, 755-764 (1982)

\bibitem[Crida and Morbidelli(2007)]{CridaMorbidelli2007} Crida, A., Morbidelli, A., \ Cavity opening by a giant planet in a protoplanetary disc and effects on planetary migration. \textit{Mon. Not. R. Astron. Soc.} 377, 1324-1336 (2007)

\bibitem[Walsh et al.(2011)]{Walsh2011} Walsh, K.~J., Morbidelli, A., Raymond, S.~N., O'Brien, D.~P., Mandell, A.~M.\ A low mass for Mars from Jupiter's early gas-driven migration.\ \textit{Nature} 475, 206-209 (2011)

\bibitem[Papaloizou and Larwood(2007)]{PapaloizouLarwood2000} Papaloizou, J.~C.~B., Larwood, J.~D.,\ On the orbital evolution and growth of 
protoplanets embedded in a gaseous disc. \textit{Mon. Not. R. Astron. Soc.} 315, 823-833 (2000)

\bibitem[Henrard(1982)]{Henrard1982} Henrard, J., apture into resonance - an extension of the use of adiabatic invariants. \textit{Cel. Mech.} 27, 3-22 (1982)

\bibitem[Murray and Dermott(1999)]{MD99} Murray, C.~D., Dermott, S.~F., \textit{Solar System Dynamics}. UK: Cambridge University Press, \ (1999)  

\bibitem[Batygin and Morbidelli(2013)]{BatyginMorbidelli2013} Batygin, K., Morbidelli, A., Analytical treatment of planetary resonances. \textit{Astron. and Astrophys.} 556, A28 (2013)

\bibitem[Yu and Tremaine(2001)]{YuTrem2001} Yu, Q., Tremaine, S.\ Resonant Capture by Inward-migrating Planets.\ \textit{Astron. J.} 121, 1736-1740 (2001)

\bibitem[Peale(1986)]{Peale1986} Peale, S.~J., Orbital resonances, unusual configurations and exotic rotation states among planetary satellites. \textit{Satellites} 159-223 Tucson: University of Arizona Press (1986)

\bibitem[Mestel(1963)]{Mestel1963} Mestel, L., \ On the galactic law of rotation.\ \textit{Mon. Not. R. Astron. Soc.} 126, 553 (1963)

\bibitem[Wisdom(1980)]{Wisdom1980} Wisdom, J., \ The 
resonance overlap criterion and the onset of stochastic behavior in the 
restricted three-body problem. \textit{Astron. J.} 85, 1122-1133 (1980)

\bibitem[Malhotra(1993)]{Malhotra1993} Malhotra, R.\ Orbital resonances in the solar nebula - Strengths and weaknesses.\ \textit{Icarus} 106, 264 (1993)

\bibitem[Landau and Lifshitz(1959)]{LandauLifshitz1959} Landau, L.~D., Lifshitz, E.~M., \textit{Fluid mechanics.\ Course of theoretical physics} Oxford: Pergamon Press (1959)

20, \bibitem[Johansen et al.(2007)]{Johansen2007} Johansen, A., Oishi, J.~S., Mac Low, M.-M., Klahr, H., Henning, T., Youdin, A.\ Rapid planetesimal formation in turbulent circumstellar disks.\ \textit{Nature} 448, 1022-1025 (2007)

\bibitem[Youdin and Goodman(2005)]{YoudinGoodman2005} Youdin, A.~N., Goodman, J., Streaming Instabilities in Protoplanetary Disks, \textit{Astrophys. J.} 620, 459-469 (2005)

\bibitem[Armitage(2010)]{Armitage2010} Armitage,  P.~J., \textit{Astrophysics of Planet Formation}, Cambridge, UK: Cambridge University Press (2010)

\bibitem[Benz and Asphaug(1999)]{BenzAsphaug1999} Benz, W., Asphaug, E.\ Catastrophic Disruptions Revisited.\ \textit{Icarus} 142, 5-20 (1999)

\bibitem[Leinhardt and Stewart(2009)]{LeinhardtStewart2009} Leinhardt, Z.~M., Stewart, S.~T.\ Full numerical simulations of catastrophic small body collisions.\ \textit{Icarus} 199, 542-559 (2009)

\bibitem[Lissauer et al.(2011)]{Lissauer2011} Lissauer, J.~J., and 38 colleagues \ A closely packed system of low-mass, low-density planets transiting Kepler-11.\ \textit{Nature} 470, 53-58 (2011)

\bibitem[Ward(1997)]{Ward1997} Ward, W.~R., Protoplanet 
Migration by Nebula Tides \textit{Icarus} 126, 261-281 (1997)

\end{thebibliography}
\end{document}